# Quantum-inspired Complex Convolutional Neural Networks

Shangshang Shi[1], Zhimin Wang[1,*], Guolong Cui[1], Shengbin Wang[1], Ruimin Shang[1], Wendong Li[1], Zhiqiang Wei[1,2], Yongjian Gu[1,*]

[1]College of Information Science and Engineering, Ocean University of China, Qingdao 266100, China

[2]High Performance Computing Center, Pilot National Laboratory for Marine Science and Technology (Qingdao), Qingdao 266100, China

*Correspondence author, e-mail: wangzhimin@ouc.edu.cn; yjgu@ouc.edu.cn

## Abstract

Quantum-inspired neural network is one of the interesting researches at the junction of the two fields of quantum computing and deep learning. Several models of quantum-inspired neurons with real parameters have been proposed, which are mainly used for three-layer feedforward neural networks. In this work, we improve the quantum-inspired neurons by exploiting the complex-valued weights which have richer representational capacity and better non-linearity. We then extend the method of implementing the quantum-inspired neurons to the convolutional operations, and naturally draw the models of quantum-inspired convolutional neural networks (QICNNs) capable of processing high-dimensional data. Five specific structures of QICNNs are discussed which are different in the way of implementing the convolutional and fully connected layers. The performance of classification accuracy of the five QICNNs are tested on the MNIST and CIFAR-10 datasets. The results show that the QICNNs can perform better in classification accuracy on MNIST dataset than the classical CNN. More learning tasks that our QICNN can outperform the classical counterparts will be found.

## 1 Introduction

In the past few years, we have witnessed many breakthroughs in quantum computing for both quantum devices [1] and quantum algorithms [2-4]. At present, there are two active directions in the research of quantum information-processing theory. One is the real quantum algorithm, which exploits the principles of quantum mechanics, i.e. quantum superposition and entanglement, and runs on quantum computers [2-4]. The other is the quantum-inspired algorithm, which borrows the quantum computing idea to improve the classical information-processing algorithm and runs on classical computers [5].

Quantum-inspired techniques have been applied in various disciplines, such as artificial intelligence, signal processing, and image processing, etc. [6,7]. Among them, quantum-inspired neural network (QINN) receives increasing amount of attention [8-18]. The first model of QINN was proposed by Kak in 1995 [11]. Subsequently, a long series of QINN models as well as learning methods have been developed based on quantum-inspired neurons [12-18]. However, all the existing QINN models are the simple three-layer feedforward networks which are unable to process high-dimensional data, and they use the quantum-inspired neurons with real-valued weights for the sake of convenience of implementing the algorithm mathematically. These situations leave two works to be done. First, the QINN is inherently a complex neural network, so it is natural to use complex values as weights in quantum-inspired neurons to construct neural networks. Second, in order to be capable of processing high-dimensional data, models of quantum-inspired deep neural networks should be developed based on the previous shallow networks. The motivation of the present work is to address the two problems.

Convolutional neural networks (CNNs) are widely used classical deep neural networks, which have been proved to be powerful in learning the complex pattern existed in the high-dimensional raw data [19-23]. They are the backbone of a majority of modern machine learning techniques.

The CNN is mainly comprised of two stages, the convolutional operations and the full-connection operations. The aforementioned quantum-inspired neurons can be applied in the full-connection stage. Furthermore, we find that the idea of quantum-inspired neurons can be employed naturally in the convolutional operations. Therefore, we extend the method of quantum-inspired neurons with complex-valued weights to the convolutional operations, and develop the models of quantum-inspired convolutional neural networks (QICNNs). Indeed, the full-connection operation is in fact the inner product of data vector and parameter vector, while the convolutional operation can be seen as the inner product of data vector (adapted from data matrix) and kernel vector (adapted from convolution kernel matrix). Specifically, we discuss five type of QICNNs with different structures in the convolutional and fully connected layers. The performance of classification accuracy of the five QICNNs are tested using the common MNIST and CIFAR-10 datasets.

It is intriguing to stress that the quantum-inspired convolutional neural networks (QICNNs) are based on complex-valued representations and operations. That is, the input real space is first mapped to the complex space in which the parameters are searched using the operations borrowed from the quantum computing. As pointed out in the recent research [24,25], using complex parameters in deep learning has numerous advantages from computational, biological and signal processing perspectives. Complex numbers have a richer representational capacity, and complex networks has the potential to enable easier optimization, better generalization characteristics, faster learning and to allow for noise-robust memory mechanisms. In the present paper, we, in fact, propose an instantiation of complex neural networks by exploiting ideas of quantum computing.

This remainder of this paper is organized as follows. In Section 2, we will describe the improved quantum-inspired neuron with complex-valued weights, and the approach to extend it to the convolutional operations. Section 3 presents the five types of QICNNs with different structures in the convolutional and fully connected layers. In Section 4, we perform a series of benchmarks of our QICNNs and demonstrate its effectiveness on standard image classification using the MNIST and CIFAR-10 datasets. Finally, conclusions are given in Section 5.

## 2 Quantum-inspired neuron and convolutional operation

### 2.1 An improved quantum-inspired neuron

In quantum computing, a qubit is implemented using a two-level quantum system. It can be described mathematically by a two-dimensional complex Hilbert space. From the quantum basic principle of state superposition, an arbitrary one-qubit state can be written as $|\varphi\rangle=\alpha|0\rangle+\beta|1\rangle$ with $\alpha$ and $\beta$ being the probability amplitude of the computational basis $|0\rangle$ and $|1\rangle$, respectively. The amplitudes $\alpha$ and $\beta$ are complex numbers and satisfy the normalization condition, i.e. $|\alpha|^2+|\beta|^2=1$. Therefore, considering the normalization constraint, the state can be rewritten as

$$|\varphi\rangle = e^{i\gamma}(\cos\theta|0\rangle + e^{i\varphi}\sin\theta|1\rangle), \tag{1}$$

where $\gamma$, $\theta$ and $\varphi$ are real numbers, and the factor $e^{i\gamma}$ and $e^{i\varphi}$ is the global and relative phase, respectively. Now we ignore the phase information in Eq. (1) and combine the magnitude information as a complex number, and then we obtain a qubit-inspired representation of input data,

$$f(\theta) = e^{i\theta} = \cos\theta + i\sin\theta. \tag{2}$$

Next, we discuss how to operate the data defined as Eq. (2). As it is well known, a single-qubit state corresponds to a point on the surface of Bloch sphere with the coordinate ($\theta$, $\varphi$). The evolution of single-qubit state can be taken as a rotation of the vector on the Bloch sphere [26]. Here, Eq. (2) corresponds to a ring on the surface of Bloch sphere, and operating the data of Eq. (2) corresponds to the rotation along latitude direction. Thus, the way of operating the data of Eq. (2) can be implemented using a similar way as quantum state evolution,

$$U\cdot f(\theta_0) = U\cdot(\cos(\theta_0)+i\sin(\theta_0)) = \cos(\theta+\theta_0)+i\sin(\theta+\theta_0). \tag{3}$$

Note that, here the operation $U$ can be itself expressed as a complex number, that is $U = f(\theta) = e^{i\theta}$, so $U\cdot f(\theta_0) = e^{i\theta}\cdot e^{i\theta_0} = e^{i(\theta+\theta_0)}$. Particularly, using $\theta' = \frac{\pi}{2}\delta - \theta$, we can define an operation similar to CNOT gate as follows,

$$f(\frac{\pi}{2}\delta-\theta) = \begin{cases} \cos\theta - i\sin\theta & (\delta=0) \\ \sin\theta + i\cos\theta & (\delta=1) \end{cases}. \tag{4}$$

When $\delta$=0, the data remains unchanged ignoring the phase information, while when $\delta$=1, the real and imaginary part interchange (corresponding to the bit flip in Eq. (1)).

So far we have defined the quantum inspired data representations and operations. Based on that a typical model of quantum-inspired neuron can be obtained as shown in Fig.1(a). In the previous works, the weights $u_i$ are taken as real numbers because it is convenient to implement in the commonly used framework of machine learning. However, the complex valued networks have richer representational capacity and better non-linearity. Therefore, we can improve the quantum-inspired neuron by directly using complex weights as shown Fig. 1(b). The output of the improved quantum-inspired neuron $S_2$ is

$$S_2 = f(\theta), \tag{5}$$

with

$$\theta = \frac{\pi}{2} - \arg(\Sigma), \tag{6}$$

$$\Sigma = \sum_{n=1}^{N} f(u_n) f(\frac{\pi}{2} x_n) - f(b). \tag{7}$$

The inputs $\{x_1, \ldots x_N\}$ are real numbers; $f(\cdot)$ represents the function as defined in Eq.(2); $\arg(\cdot)$ is to calculate the argument of a complex number. We perform several numerical experiments and find that the fully connected network with complex-valued weights indeed performs better than that with real weights. So we will employ the improved quantum-inspired neuron to construct the convolutional neural networks.

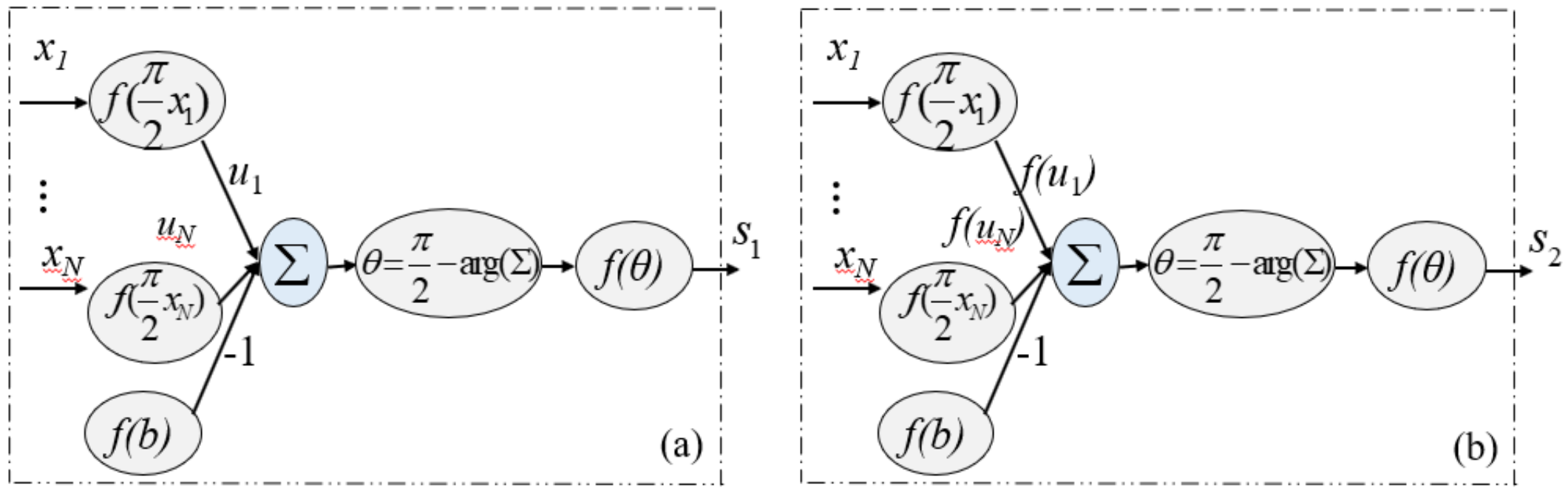

Fig. 1 The structure of the common quantum-inspired neuron with real valued weights (a), and the improved quantum-inspired neuron with complex weights (b). The weights $\{u_1, u_2 \ldots u_N\}$ are real numbers. The module Σ represents weighted summation.

## 2.2 Quantum-inspired convolutional operations

In general, the convolutional neural network (CNN) is comprised of two main parts, namely the convolutional operations and the full-connection operations. It is evident that the improved quantum-inspired neuron can be applied to implement the full-connection stage. Moreover, a similar way can also be used to perform the convolutional operation as depicted in Fig. 2. The input pixel and the convolutional kernel are first transformed into complex space, and then the multiplication between them is just like the process of weighted summation as shown in Fig. 1(b). After the convolutional operation (namely a rotation operation as shown in Eq. (3), a NOT operation is added as it is in the quantum-inspired neurons.

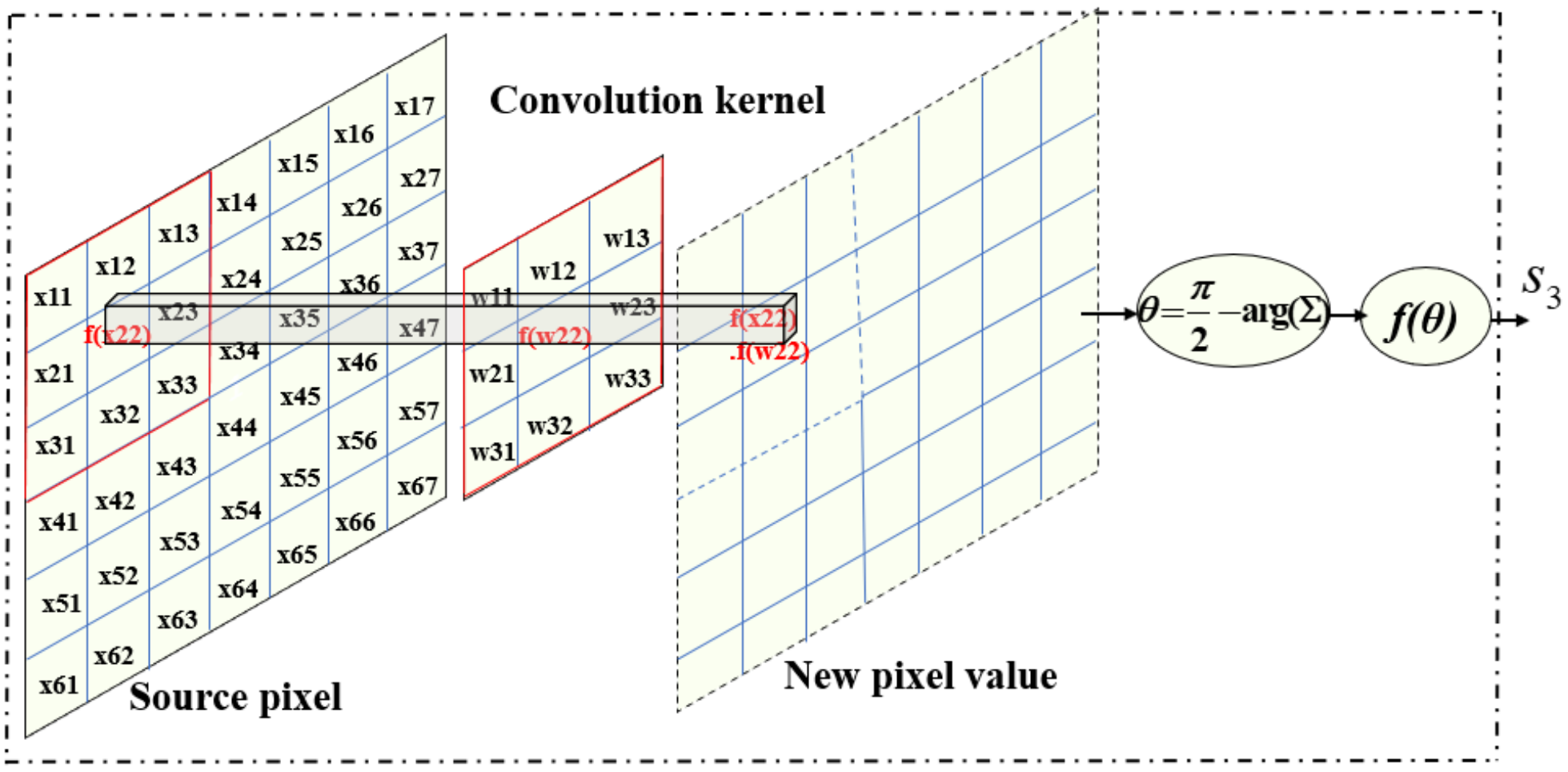

Fig. 2 The quantum-inspired method of implementing the convolution operations in convolutional neural networks.

Now that we have developed the models of quantum-inspired neutron and quantum-inspired convolutional operation, we can naturally construct the quantum-inspired convolutional neural networks (QICNNs) by combining them. However, since most present machine learning frameworks are based on real-valued representations and operations, our complex-valued neutron and convolutional operations should be first adapted to be able to implemented conveniently on the common frameworks.

As shown in Eq. (2) and (3), both the data and operation can be expressed as a complex number. A complex number $z=a+ib$ has a real component $a$ and an imaginary component $b$. The basic idea is that the real part and imaginary part are represented as logically distinct real valued entities and simulate complex arithmetic using real-valued arithmetic internally [24]. For example, given a typical real-valued input data with $N$ feature maps, we transform them to complex values as shown in Fig. 1. Then the $N$ feature maps are expanded to $2N$, of which the first $N$ feature maps are allocated to represent the real components and the remaining $N$ to represent the imaginary ones.

Next, the convolutional operations on the complex values are based on the multiplication principle of complex numbers. Specifically, the data can be expressed as $x = x_{real} + ix_{imag}$ with $x_{real}$ and $x_{imag}$ being the real matrices of two feature maps, while the complex filter matrix is written as $W = w_{real} + iw_{imag}$ with $w_{real}$ and $w_{imag}$ being the real matrices to be learned. Then the convolutional operation is implemented as follows:

$$W * x = (w_{real} * x_{real} - w_{imag} * x_{imag}) + i\,(w_{imag} * x_{real} + w_{real} * x_{imag}). \quad (8)$$

Using this equation in Fig. 2, the quantum-inspired convolutional operation is realized in the way as shown in Fig. 3.

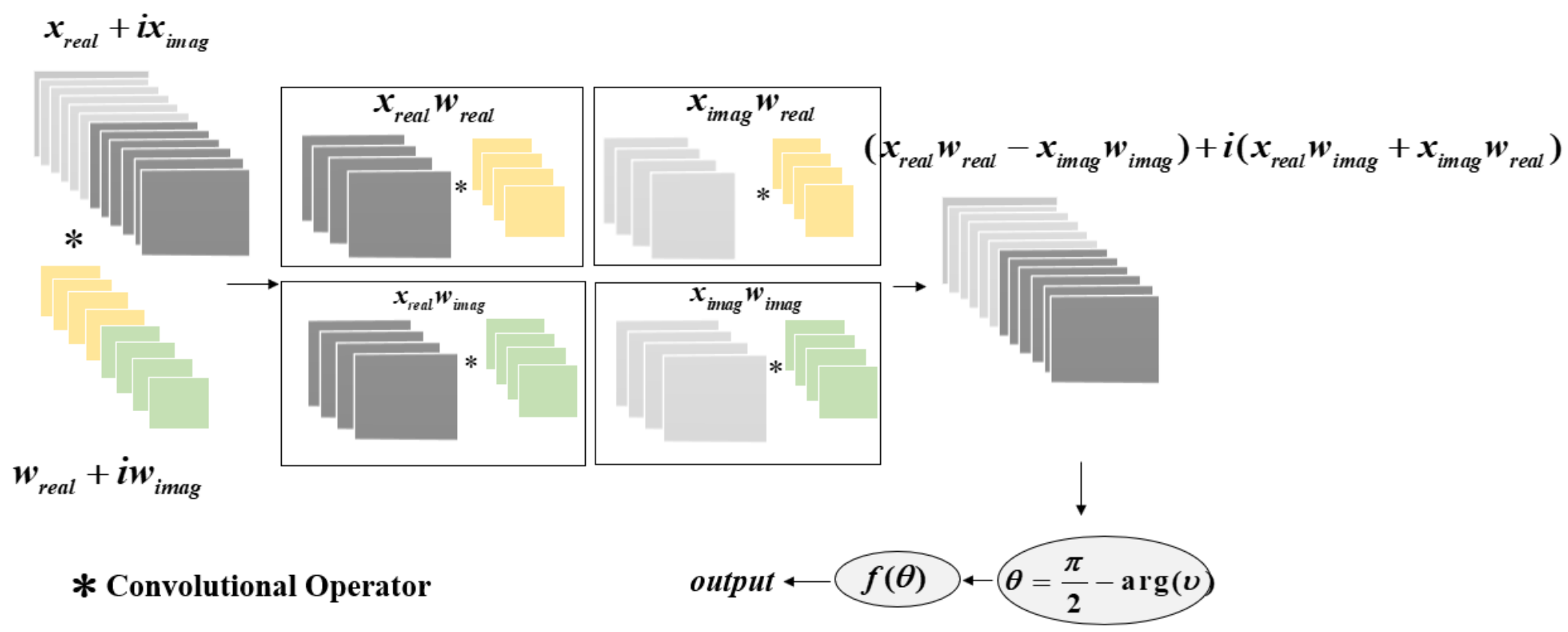

Fig. 3 The practical implementation of the quantum-inspired convolution operations.

Specifically, the quantum-inspired convolutional operation is implemented through the following five steps:

1st, Transform the input and filter data to complex space using Eq. (2), i.e. $x_0 \mapsto f(x_0)$ $w_0 \mapsto f(w_0)$. That is, $x = x_{real} + ix_{imag}$, and $w = w_{real} + iw_{imag}$.

2nd, Perform the complex convolutional operation between complex values, i.e. $x * w = (x_{real} * w_{real} - x_{imag} * w_{imag}) + i(x_{real} * w_{imag} + x_{imag} * w_{real})$.

3rd, Calculate the phase, i.e. $arg(\upsilon) = \arctan(\frac{x_{real} * w_{imag} + x_{imag} * w_{real}}{x_{real} * w_{real} - x_{imag} * w_{imag}})$.

4th, Perform an NOT operation, i.e. $\theta = \frac{\pi}{2} - \arg(\upsilon)$.

5th, Compute the output result, i.e. $output = |\mathrm{Im}(y)|^2$, $y = f(\theta)$.

Note that the second step represents the rotation operation on each input data and the fourth step is a global NOT operation on the sum of input data.

## 3 The architecture of quantum-inspired convolutional neural networks

In order to show clearly the architecture of QICNNs, we take a classical CNN[29] as the template to build the QICNNs. The classical template CNN is adapted from the LeNet-5 model which is widely used for handwritten digit and English letter recognition. In general, the template CNN consists of three convolutional layers and two fully connected layers. Fig. 4(a) shows the implementing details of the template CNN.

The straightforward way to build the QICNNs is to employ the improved quantum-inspired neuron (see Fig. 1(b)) in the fully connected layers and the quantum-inspired convolutional operation (see Fig. 3) in the convolutional layer. Furthermore, we find that using the quantum-inspired techniques in different stages of convolutional and fully connected layers will have a great influence on the performance of the network. Specifically, we study the five cases of using the quantum-inspired techniques in the template CNN, and provide five types of QICNN as shown in Fig. 4(c).

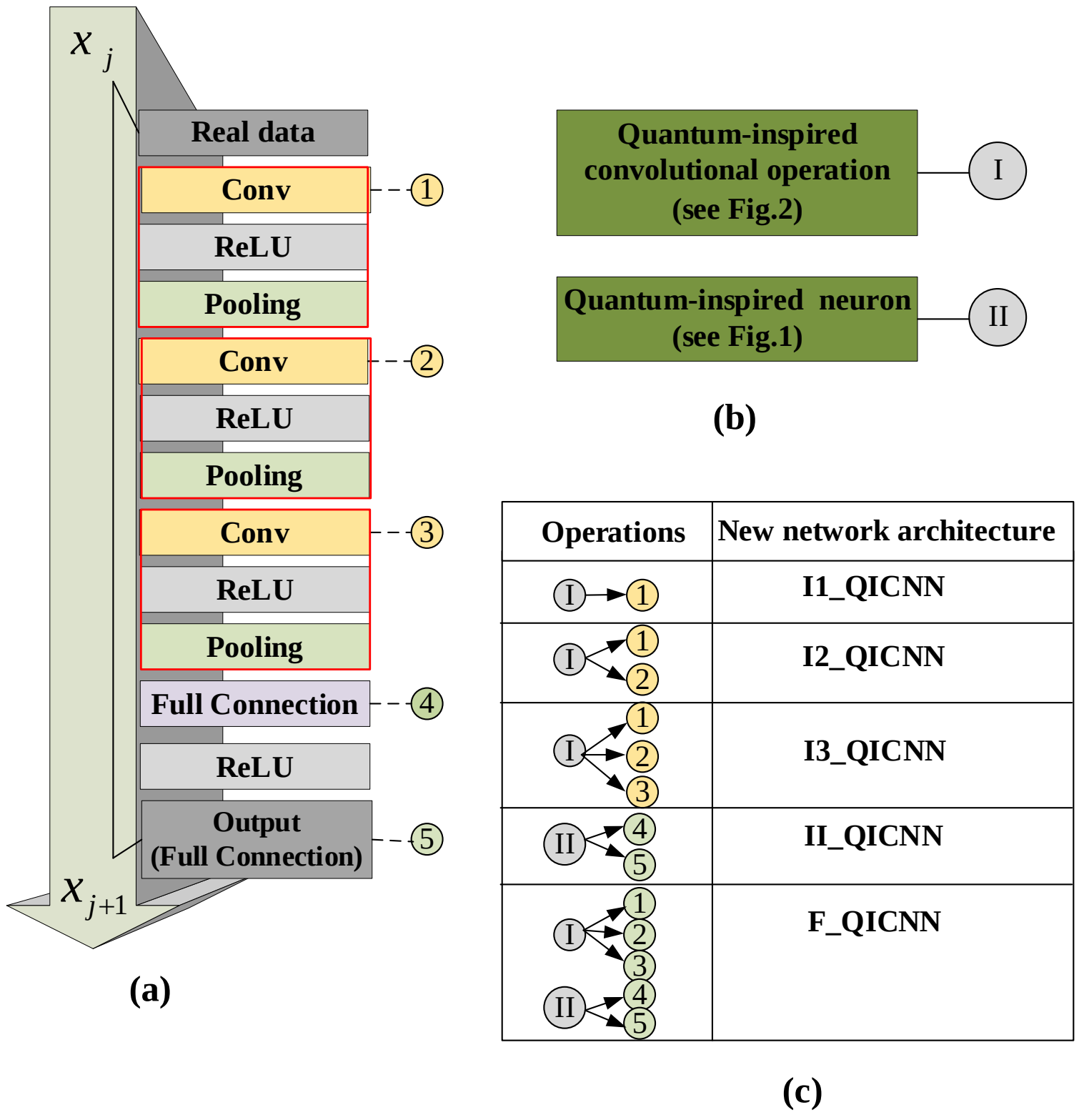

Fig. 4 The strategy to build QICNNs based on the classical template CNN. (a) The implementing details of the template CNN; (b) the two building blocks of the QICNN, namely the quantum-inspired convolutional operation (I) and the improved quantum-inspired neuron (II); (c) five ways to build QICNNs as discussed in the text.

As shown in Fig. 4, the first QICNN is called I1_QICNN, for its first convolutional layer uses the quantum-inspired convolutional operation and the others are the classical ones. Accordingly, it is easy to understand the structures of I2_QICNN and I3_QICNN. The first three models are also called I_QICNN. The fourth one, II_QICNN, uses the improved quantum-inspired neuron in the fully connected layers. The last one called as F_QICNN is the full quantum case. Below we present the mathematical details of implementing QICNNs.

### 3.1 Mathematical details of implementing I_QICNN

We take the model of I3_QICNN as an example to describe the mathematical details of the I_QICNN algorithms.

(1) Input layer. In this layer, the input matrix is rescaled in the range [0, 1], and multiply $\pi/2$ to get the phase of the complex input in the range [0, $\pi/2$]. The output of the input layer can be calculated as:

$$x_{input1} = \frac{\pi}{2} x_{data} \tag{9}$$

$$x_{real} = \cos(x_{input1}),\ x_{imag} = \sin(x_{input1}) \tag{10}$$

(2) Convolutional layer 1. This is the first convolutional layer in which a quantum-inspired convolutional operation is used. The real and imaginary part of the convolutional kernel matrix are obtained by $w_{real1} = \cos(w_0),\ w_{imag1} = \sin(w_0)$. It is the same for the bias term, i.e. $b_{real1} = \cos(b_0), b_{imag1} = \sin(b_0)$. Then the convolutional operation is performed as follows,

$$\begin{aligned} y_{c1_real} &= conv(x_{real}, w_{real1}) - conv(x_{imag}, w_{imag1}) - b_{real1} \\ y_{c1_imag} &= conv(x_{imag}, w_{real1}) + conv(x_{real}, w_{imag1}) - b_{imag1} \end{aligned} . \quad (11)$$

In order to perform the controlled NOT operation as shown in Fig.3, we calculate the angle using $\arg_1(\upsilon) = \arctan(\frac{y_{c1_imag}}{y_{c1_real}})$. The output is $y_{c_1} = f(\frac{\pi}{2} - \arg_1(\upsilon))$.

(3) ReLu layer 1. The ReLu function is used as the activation function, i.e. $y_{relu1_real} = \mathrm{Re}LU(y_{c_1_real}), y_{relu1_imag} = \mathrm{Re}LU(y_{c_1_imag})$.

(4) Pooling layer 1. The pooling is the down-sampling (sub-sampling) process. It can reduce the amount of data processing while retaining useful information. Maximum pooling is used here, namely $y_{pool1_real} = \max pool(y_{relu1_real})$ and $y_{pool1_imag} = \max pool(y_{relu1_imag})$.

(5~7) The second convolutional layer. The above steps of (2), (3) and (4) are repeated for Convolutional layer 2, ReLu layer 2 and Pooling layer 2.

(8~10) The third convolutional layer. Since the following fully connected layer after the third convolutional layer is based on real values, the imaginary part of the output of Convolutional layer 3 is remained as the input of ReLu layer 3. That is, after the complex value convolutional operation as Eq. (6), the angle is $\arg_3(\upsilon) = \arctan(\frac{y_{c3_imag}}{y_{c3_real}})$, then the output is $y_{c_3} = |\mathrm{Im}(f(\frac{\pi}{2} - \arg_3(\upsilon)))|^2$. The operations for ReLu layer 3 and Pooling layer 3 are the same as above.

(11) Fully connected layer 1. The role of the fully connected layer is mainly to achieve classification. This layer uses the real-valued classical neuron. First, the weighted summation is $y_{f1} = y_{pool3} * w_{f1} + b_{f1}$, where $y_{pool3}$ is the output of the Pooling layer 3; $w_{f1}$ is the connection weights between convolutional layer 3 and fully connected layer 1; $b_{f1}$ is the threshold of the fully connected layer 1. Then the action of ReLu function is $y_{f1_relu} = relu(y_{f1})$.

(12) Output layer. The output of the network is $y_{output} = y_{f1_relu} * w_0 + b_0$ with $w_0$ being the weight of the output layer and $b_0$ being the threshold of the output layer.

### 3.2 Mathematical details of implementing II_QICNN

The type of II_QICNN is the network using classical convolutional operation in the convolutional layers and the improved quantum-inspired neurons in the fully connected layers. For the II_QICNN, the first three convolutional-ReLu-pooling operations are the same as the classical ones; the implementing procedures of the fully connected layers is implemented as follows.

(1) Input layer. The input data is transformed into complex space with phase values

in the range [0, $\pi$/2]. That is $x_{input} = \frac{\pi}{2} y_{pool3}$ and $y_{input} = f(x_{input})$ with $y_{pool3}$ being the output of the third Pooling operation.

(2) Fully connected layer 1. According to Eqs. (5)-(7), the output of this layer is calculated as $\upsilon_1 = f(w_1) y_{input} - f(b_1)$ , $\theta_1 = \frac{\pi}{2} - \arg(\upsilon_1)$ and $y_{output1} = f(\theta_1)$ . The parameter $w_1$ is the connection weight matrix between the Pooling layer and the fully connected layer 1, and $b_1$ is the threshold vector of the fully connected layer 1.

(4) Fully connected layer 2. This layer is the output layer. The output is slightly different from that of fully connected layer 1. The output is calculated by $\upsilon_2 = f(w_2) y_{output1} - f(b_2)$ , $\theta_2 = \frac{\pi}{2} - \arg(\upsilon_2)$ , $y_{output2} = f(\theta_2)$ and $y_{output} = |\operatorname{Im}(y_{output2})|^2$ . The parameter $w_2$ is the connection weight matrix between fully connected layer 1 and fully connected layer 2; $b_1$ is the threshold vector of the fully connected layer 2; $y_{output}$ is the final output of II_QICNN.

**4 Experimental results and discussions**

In order to test the efficiency of the present QICNNs, we use the five types of QICNNs to perform the image recognition on the MNIST and CIFAR-10 dataset. Below, we show the performance of convergence rate and classification accuracy of the five types of QICNNs.

MNIST is a dataset used for image recognition of handwritten digits [30]. It contains handwritten digits of values ranging from 0 to 9. CIFAR-10 is a dataset used for object detection, which is labeled a subset of 80 million tiny images [31]. Both MNIST and CIFAR-10 dataset contain 50,000 training images and 10,000 testing images.

The performance of loss of the five QICNNs as well as the classical template CNN are shown in Fig.5. It shows that the loss performance of QICNNs depends on both the structure of QICNNs and the dataset processed. For the MINIST dataset, the structure of II_QICNN has the fastest rate of convergence, which is remarkably better than the template CNN. However, for the CIFAR-10 dataset, the performance of template CNN is slightly better than II_QICNN. Therefore, the QICNNs can achieve acceleration for specific problems. In general, employing the improved quantum-inspired neurons in the fully connected layers performs better than employing the quantum-inspired convolution operation in the convolution layers.

Fig.6 shows the classification accuracy of the five QICNNs and the classical template CNN on the MNIST and CIFAR-10 dataset running in a single CPU unit. For the MINIST dataset, the model of II_QICNN has the maximum classification accuracy (about 0.9965), which is remarkably better than the template CNN (about 0.9950). For the CIFAR-10 dataset, the accuracy of the template CNN is slightly better than II_QICNN. As before, employing the improved quantum-inspired neurons in the fully connected layers performs better than employing the quantum-inspired convolutional operations in the convolutional layers. In addition, by comparing the classification accuracy of I_QICNN as shown in Fig.6, it seems that the more quantum-inspired convolutional operations are used, the lower the accuracy is.

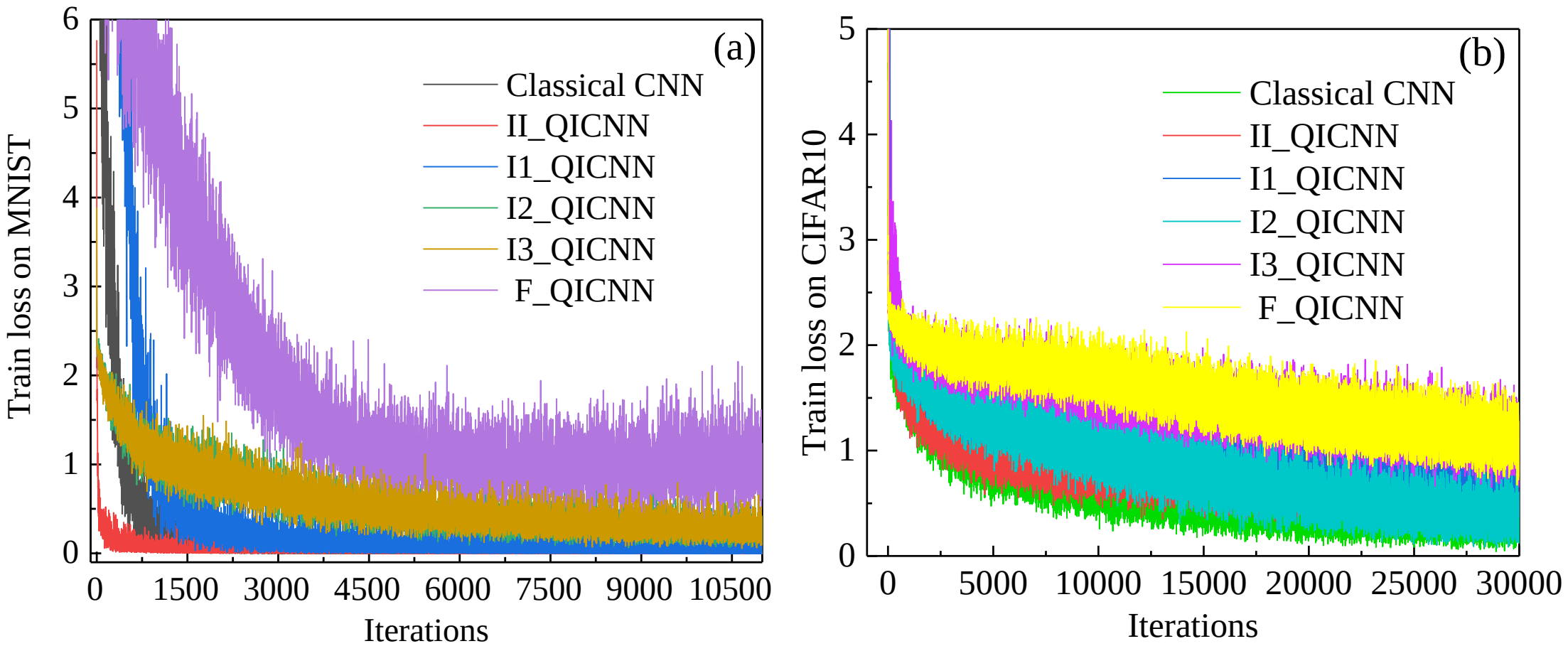

Fig.5 The experimental results of loss curve for the five types of QICNNs and the classical template CNN performed on the dataset of MNIST (a) and CIFAR-10 (b).

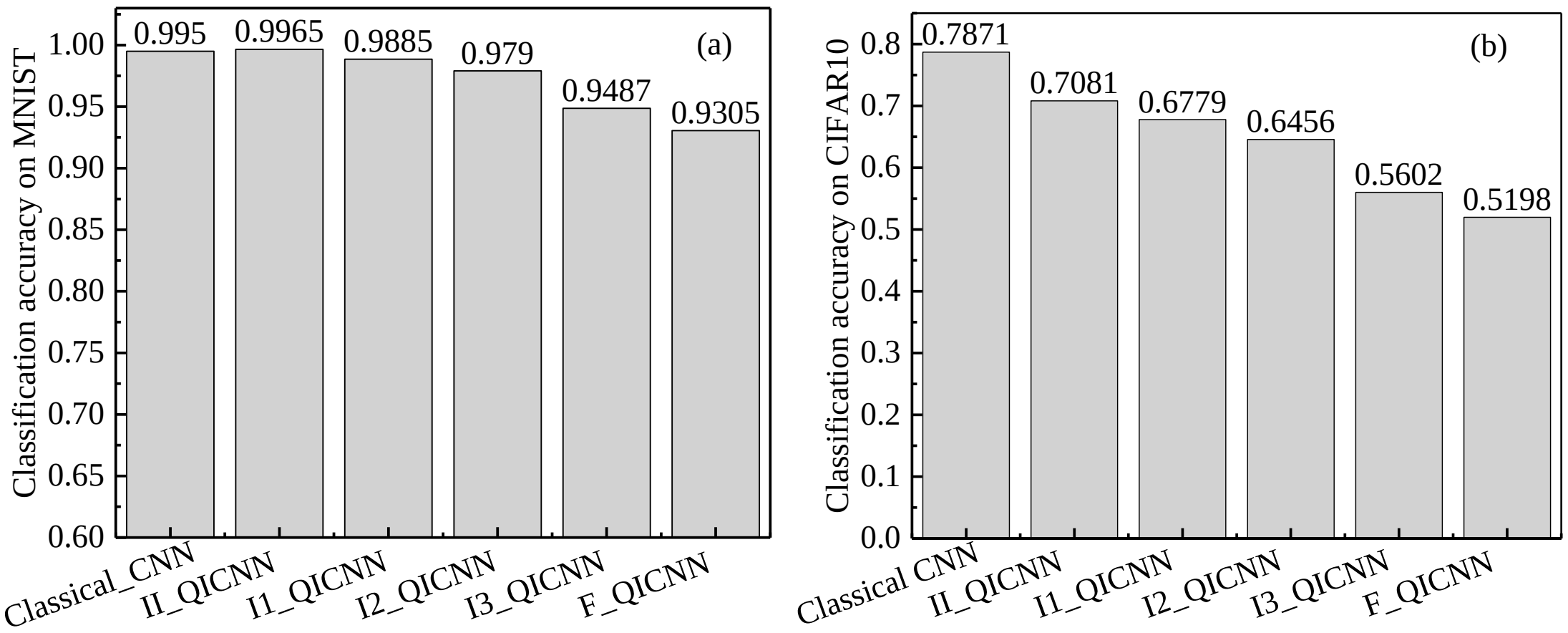

Fig.6 The experimental results of accuracy for the five types of QICNNs and the classical template CNN performed on the dataset of MNIST (a) and CIFAR-10 (b).

## 5 Conclusions

In the present work, we improve the quantum-inspired neurons by exploiting the complex-valued weights to have richer representational capacity and better non-linearity. Then the basic idea is extended naturally to perform the convolutional operations. By employing the quantum-inspired neurons in the fully connected layers and/or the convolutional operations in the convolutional layers, we develop five types of quantum-inspired convolutional neuron networks (QICNNs). The mathematical details of implementing the QICNNs are developed, which can be executed based on the common real-valued framework of machine learning.

The performance of the five types of QICNNs on image recognition is studied using the MNIST and CIFAR-10 dataset. The results show that the QICNN using the improved quantum-inspired neuron in the fully connected layer, i.e. II_QICNN has the best performance. Its classification accuracy on the MNIST dataset is remarkably higher than the classical CNN.

In the future, we will further optimize the QICNNs by improving the method of implementing the complex-valued deep neural networks. In addition, we will find more specific datasets and learning tasks that the QICNN can outperform the classical CNNs.

**Acknowledgements**
The present work is supported by the National Natural Science Foundation of China (Grant No. 12005212, 61575180) and the Pilot National Laboratory for Marine Science and Technology (Qingdao).

**Declarations**

**Conflict of interest** The authors declare that they have no known competing financial interests or personal relationships that could have appeared to influence the work reported in this paper.

**Code availability** The codes are available upon request from the authors.